\documentclass[12pt]{iopart}

\usepackage[utf8]{inputenc}	
\usepackage{graphicx}		
\usepackage{dcolumn}		
\usepackage{bm}			
\usepackage{amssymb}
\usepackage{subscript}
\usepackage{textcomp}		
\usepackage{color}		
\usepackage{lipsum}
\usepackage{epsfig}		
\usepackage{subfigure}

\begin{document}

\title[]{Influence of non-collisional laser heating on the electron dynamics in dielectric materials}

\author{L. Barilleau, G. Duchateau, B. Chimier, G. Geoffroy, \\ and V. Tikhonchuk}

\address{Université de Bordeaux-CNRS-CEA, Centre Lasers Intenses et Applications, UMR 5107, 351 Cours de la Libération, 33405 Talence, France}
\ead{guillaume.duchateau@u-bordeaux.fr}
\vspace{10pt}

\begin{abstract}
The electron dynamics in dielectric materials induced by intense femtosecond laser pulses is theoretically addressed. The laser driven temporal evolution of the energy distribution of electrons in the conduction band is described by a kinetic Boltzmann equation. In addition to the collisional processes for energy transfer such as electron-phonon-photon and electron-electron interactions, a non-collisional process for photon absorption in the conduction band is included. It relies on direct transitions between sub-bands of the conduction band through multiphoton absorption. This mechanism is shown to significantly contribute to the laser heating of conduction electrons for large enough laser intensities. It also increases the time required for the electron distribution to reach the equilibrium state as described by the Fermi-Dirac statistics. Quantitative results are provided for quartz irradiated by a femtosecond laser pulse with a wavelength of 800 nm and for intensities in the range of tens of TW/cm$^2$, lower than the ablation threshold. The change in the energy deposition induced by this non-collisional heating process is expected to have a significant influence on the laser processing of dielectric materials.
\end{abstract}

%
%
%
%
%

\section{INTRODUCTION}
\label{section1}
Short and intense laser pulses are widely used for material structuration, including metals~\cite{PhysRevB.65.214303, PhysRevB.87.035414}, semi-conductors~\cite{PhysRevB.54.4660} and dielectrics~\cite{PhysRevB.61.11437, Appl.Phys.A.110.579}. Large band gap dielectric materials (such as silica, quartz, sapphire, etc) are used for various applications going from laser structuration of materials (waveguides, gratings, etc) to new technologies for medicine \cite{Bulgakova2006, Bulgakova2010, Rep.Prog.Phys.76.036502}. The local changes in the material properties resulting from complex laser-matter interactions may be described as follows. First, the valence electrons are promoted to the conduction band (CB) through photons absorption or tunneling. The conduction electrons can then further absorb the laser energy to be driven to higher energy levels. In the same time, they undergo collisions with phonons or other electrons leading to their relaxation and the lattice heating on longer timescales. For laser intensities close to the material breakdown threshold or for non linear optical materials, the electron dynamics may in turn affect the pulse propagation, leading to a strong coupling between both the laser pulse propagation and the electron dynamics \cite{PhysRevE.85.56403, DuchateauBourgeade, GulleySPIE, Gulley, DuchateauPop, Varin}. In all cases, the absorption of the laser pulse energy by the material, which the control is crucial for various experiments and applications, is directly related to the laser driven electron dynamics. Understanding the fundamental physical processes driving the electron dynamics is thus a key issue to make accurate predictions for the laser energy deposition.

In the case of femtosecond laser pulses, the interaction time is much shorter than the characteristic time of the electron-phonon coupling which is of the order of a few picoseconds. A significant energy transfer from the electrons to the lattice takes place for times significantly longer than the pulse duration. The electron dynamics on the pulse timescale is thus mainly decorrelated from any significant lattice evolution. Various approaches allow one to describe the electron dynamics in dielectric materials on such timescales. The \textit{ab initio} calculations based on the time-dependent density functional theory (TDDFT) can describe the electron interaction with the laser electric field \cite{Otobe2008, Otobe2010}. However, within this approach, the electron-electron interactions are not described properly \cite{PhysRevB.77.165104}. In addition, since TDDFT is very CPU-time consuming, its coupling to the Maxwell's equations to describe the coupled laser pulse propagation and electron dynamics is not conceivable nowadays. Another approach, less cumbersome, is based on the kinetic Boltzmann's equation \cite{PhysRevB.61.11437, Appl.Phys.A.110.579, J.Opt.Soc.Am.B.31.C28}. It describes Markovian interactions between electrons, photons, ions, and phonons in the bulk material. This approach renders it possible to describe the evolution of the electron energy distribution including all the possible collisional processes. Since the contribution of various interactions may be independently analysed, this model offers an efficient approach to understand the laser induced electron dynamics in dielectric materials.

Such an approach has been developed for studying the electron dynamics induced by intense femtosecond laser pulses in dielectrics \cite{PhysRevB.61.11437} and metals \cite{PhysRevB.65.214303}. The temporal evolution of the electron energy distribution due to various collisional processes and characteristic relaxation times have been obtained. In particular, the electron heating has been described by the ion or phonon-assisted absorption of photons in the CB \cite{Appl.Phys.A.110.579}. However, experimental and theoretical investigations have shown that photon absorption in the CB through {\it {non-collisional}} processes may play an important role for the electron dynamics in solids \cite{Appl.Phys.A.98.679, PhysRevB.74.235215, J.PhysiqueIV.108.113, Europhys.Lett.67.301, Hawkins2015, Bulgakova2004, Bulgakova2016}. Indeed, higher energies of photo-emitted electrons than expected with standard collisional processes were observed. The non-collisional process relies on direct transitions between sub-bands of the CB through multiphoton absorption, hereafter referred to as the multiphoton interband process (MIP). However, an accurate description of the electron dynamics, including the MIP, is not yet available.

The first proposed expression for the MIP rate only includes the transitions between the bottom of the conduction band to the first excited sub-band \cite{Appl.Phys.A.98.679}. In Section \ref{section2} of the present work, the expression for the MIP rate is revisited and generalized for transitions to various sub-bands with higher energies. This Section also introduces this non-collisional heating mechanism into the Boltzmann kinetic equation in addition to all standard collisional interactions for electron excitation and relaxation. Predictions are made in Section \ref{section3} for physical conditions of wide interest for many applications \cite{Rep.Prog.Phys.76.036502}: interaction of femtosecond laser pulses with intensities in the range of tens of TW/cm$^2$ at the wavelength of 800 nm (corresponding to the Ti:Saphir laser) with quartz (crystalline phase of SiO$_2$). It is shown that the interband process has a significant influence on the electron dynamics, confirming the above-mentioned experimental observations: the electrons can reach higher kinetic energies in the course of laser interaction. This also leads to an increase in the time required for the electrons to reach the thermal equilibrium as described by the Fermi-Dirac statistics. A summary of the present work and outlooks are drawn in Section \ref{section4}.

\section{Theoretical model}
\label{section2}
\subsection{Band structure and general model}
The description of the electron dynamics including various collisions and the MIP first requires the knowledge of the band structure of the studied material. The most accurate description of the CB structure requires \textit{ab initio} calculations based on the density functional theory. While such an approach may describe the MIP \cite{Otobe2008, Otobe2010, Mezel}, it is technically difficult to couple it to other collisional processes. The framework of a kinetic approach is better suited for description of such a coupling. It has been shown that the description of the CB by multiple parabolic bands provides correct estimations of the MIP rate \cite{J.PhysiqueIV.108.113, Europhys.Lett.67.301, Appl.Phys.B.78.989, Phys.Status.Solidi.C.2.240, Hawkins2015}. As described in the following Section \ref{MPinterband}, each parabolic sub-band then corresponds to a multiple of the reciprocal lattice wavevector $\vec{G}$.

The modeling of the electron dynamics is performed as follows. First, the multiple parabolic band structure is used to evaluate the multiphoton interband rate which provides the density of electrons (in the bottom of the CB) promoted to an excited energy $E_k$ per unit time. The MIP rate, called $\partial w_{1f}/\partial p$, is derived in Section \ref{MPinterband}. For the sake of simplicity, in order to describe the whole electron dynamics through the kinetic Boltzmann equation as developed in \cite{PhysRevB.61.11437}, only one parabolic sub-band ($\vec{G}=\vec{0}$) is considered in the description of collisional operators. This assumption is relative to the removing of umklapp processes involving transitions assisted with the lattice wavevector. The latter process has been shown to increase the laser absorption by roughly 20 \% \cite{Shcheblanov}. The multiphoton interband rate is then introduced in the kinetic model as an usual collisional operator only depending on the considered final electron energy, the details of the required more complex band structure being encapsulated. This procedure is similar to the one of introducing multiphoton ionization in the kinetic modeling \cite{PhysRevB.61.11437}. The details of this procedure are provided in Section \ref{kinetic}.

\subsection{Rate of the multiphoton interband process}
\label{MPinterband}
The MIP accounts for direct transitions between electronic sub-bands of the CB through multiphoton absorption. The model proposed in \cite{J.PhysiqueIV.108.113, Europhys.Lett.67.301} accounts only for the transition between the first and the second band, limiting the highest energy that may be reached by electrons through this process. Here this model is generalized by accounting for transitions to upper bands. Transitions with higher multiphoton orders are then allowed, which are expected for sufficiently high laser intensities, typically in the range of tens of TW/cm$^2$.

\begin{figure}[t] 
  \begin{center}
    \includegraphics[width = 10cm, trim = 0cm 0cm 0cm 3cm, clip] {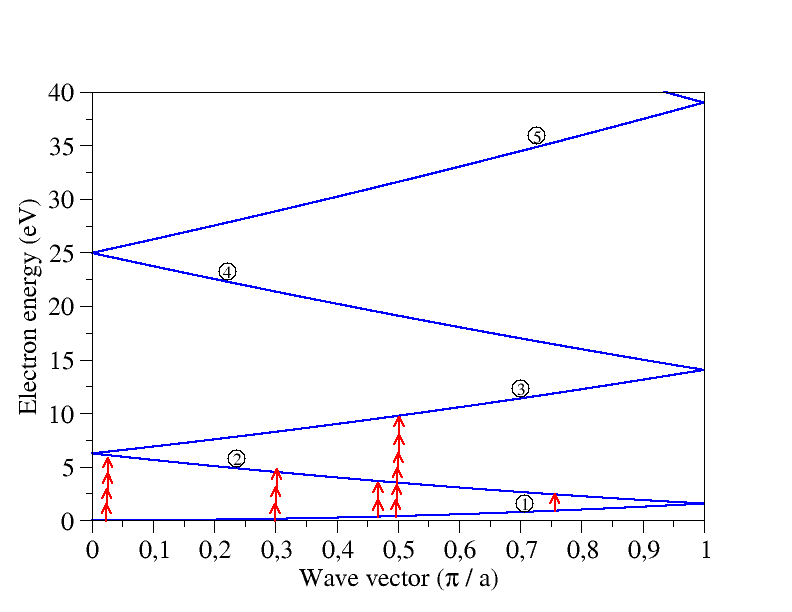}
  \end{center}
  \caption{Structure of the conduction band in the first Brillouin zone as described in the multiple parabolic band model. The lattice period is set to 4.91 \AA. An illustration of possible multiphoton transitions is depicted by the arrows. Depending on both the multiphoton order and the final sub-band, the wave vector is different according to energy conservation considerations.} 
  \label{image_Structure_band}
\end{figure}

The proposed model for the interband multiphoton rate is then as follows. First, the description of the various sub-bands in the CB is obtained by considering one active electron in a one-dimensional periodic potential. A simple cubic structure for the Bravais lattice is used in order to further simplify the calculations. Under this framework, the energy $E_{k}^{(b)}$ of the band $b$ in the first Brillouin zone depends on a multiple $j(b)$ of the reciprocal lattice wave vector $\vec{G} = 2\pi/a$, where $a$ is the size of the lattice cell in the real space. This energy reads \cite{Ashcroft}:
\begin{equation}
  E_{k}^{(b)} = \frac{\hbar^2 \left( \vec{k} - j(b) \vec{G} \right)^2}{2 m}
\end{equation}
where $\vec{k}$ is the electron momentum and $m$ is the effective electron mass in the CB. The value of $j(b)$ is such that j(1) = 0, j(2) = 1, j(3) = -1, j(4) = 2, j(5) = -2, etc \cite{Ashcroft}. An illustration of such a band structure for the CB is provided by Fig. \ref{image_Structure_band} where the five lowest bands of the first Brillouin zone are shown. The first band, $b=1$, is in the bottom of the CB. The electrons may be excited to higher sub-bands, with $b \ge 2$, through multiphoton absorption. Note that only the first sub-band is considered as an initial state since valence electrons are mainly promoted to the bottom of the CB during the ionization process.

The rate per unit volume, $w_{1f}$, for electron transitions from the lowest band ($b=1$) to a higher band $b=f$ (with $f\geq 2$) through multiphoton absorption may then be derived following a formalism as developed in \cite{JETP.20.1307, Europhys.Lett.67.301, Gruzdev}. It is based on the evaluation of the quantum transition amplitude where Volkov states \cite{Volkov} are used to describe the multiphoton absorption. It is worth noting that intermediate states between initial and final sub-bands are not taken into here, neither intraband transitions \cite{Ridley, Hawkins2015}, their roles should be to further increase the influence of this process on the electron dynamics since the transition amplitude is expected to increase in that case \cite{DuchateauThese, Rodriguez, Catoire}. As shown in Appendix \ref{app:interband}, the MIP rate per unit momentum, which will be used in the kinetic approach, reads:
\begin{equation}
  \frac{\partial w_{1f}}{\partial p}  = \frac{m}{2 \pi \hbar^4 \left| p p_f \right|} V_{1f}^2 I \sum\limits_{n} J'^2_n(B_{1f}) \Theta (p-p_n)
  \label{dWif}
\end{equation}
where $p=\hbar k$, $\Theta$ is the Heaviside function, and the other quantities are defined hereafter. As shown in Appendix \ref{app:interband}, after integration over the momentum amplitude, the total MIP rate reads:
\begin{equation}
  w_{1f}  = \frac{m}{4 \pi \hbar^2 \left| p_f \right|} V_{1f}^2 I \sum\limits_{n} J'^2_n(B_{1f}) \left[ \left( \frac{\pi}{a} \right)^2 - k_n^2 \right]
  \label{Wif}
\end{equation}
where $I$ is the laser intensity, $V_{1f}$ is proportional to the dipolar matrix element, and $n$ is the number of photons bridging the initial bottom and final excited sub-bands of the CB. There are several allowed values of $n$ which account for the various allowed transition pathways depending on the electron wave vector $k_n$ in the Brillouin zone as illustrated in Fig. \ref{image_Structure_band}. The value of $k_n$ ($p_n = \hbar k_n$) is provided in Appendix \ref{app:interband}. $J'_n(B_{1f})$ is the derivative of the Bessel function with the argument:
\begin{equation}
  B_{1f} = \frac{1}{\hbar \omega} \frac{e \vec{F} \left( t \right) . \left( \vec{p}_f - \vec{p}_1 \right)}{m \omega}
\label{eq:B1f}
\end{equation}
where $\hbar \omega$ is the photon energy, $\vec{F}(t)$ is the envelope of the laser electric field, and $\vec{p}_b = \hbar j(b) \vec {G}$. For a transition to the first excited sub-band $b=2$, our expression of the MIP rate (\ref{Wif}) is exactly the same as the one provided in \cite{Europhys.Lett.67.301} before integration over the momentum amplitude. For transitions to higher sub-bands, the mathematical structure of the presently derived rate remains similar but the number of photons to bridge the initial and final sub-bands is larger.

In order to evaluate the matrix element $V_{1f}$, numerical calculations solving a one-dimensional time-dependent Schrödinger equation (TDSE) have been performed. Within this approach, one-active electron in a periodic pseudopotential is considered, and the wavefunction is expanded over a plane wave basis set. More details of this approach can be found in \cite{PhysRevB.74.235215}. With a lattice periodicity of $4.91 \AA$, a comparison of predictions of the analytical expression (\ref{Wif}) with TDSE calculations indicate that $V_{1f}^2$ is of the order of $9 \times 10^{-55}$ J.m\textsuperscript{2}.s within the range of intensities used in this work, which is consistent with the value reported in \cite{Europhys.Lett.67.301}. The order of magnitude of this value may also be found with simple considerations based on approximate wavefunctions.


\subsection{Kinetic equation}
\label{kinetic}
The energy distribution $f(E_k,t)$ of electrons in the CB is evaluated by solving the Boltzmann kinetic equation \cite{PhysRevB.61.11437, Appl.Phys.A.110.579}. In order to simplify this problem, first the laser field is considered as homogeneous in space so that transport processes are neglected. Second, external forces are not taken into account. Indeed, this study addresses interactions below the ablation threshold where the produced electron density in the CB remains small compared to the critical plasma density. Both the whole material and the electron system are thus not significantly perturbed by laser irradiation. Third, the electron system is assumed to be isotropic so that it may be described by its energy distribution. Then, the temporal evolution of the electron energy distribution function is given by:
\begin{equation}
  \frac{\partial}{\partial t} f(E_k, t)  =  \left. \frac{\partial f(E_k, t)}{\partial t} \right|_{{{I}}}
  + \left. \frac{\partial f(E_k, t)}{\partial t} \right|_{{{Heat}}}
  + \left. \frac{\partial f(E_k, t)}{\partial t} \right|_{{{Relax}}}
  - \frac{f(E_k, t)}{\tau_{r}}
  \label{eq:Boltzmann}
\end{equation}
where the various terms of the right hand side correspond to the so-called collisional operators which account for various physical processes of energy exchange between particles. It is worth noting that these operators account for total energy and momentum conservation of colliding particles. The first term of the right side corresponds to the photo-ionization and impact ionization contribution, i.e. an electron transition from the valence band (VB) to the CB. The second term of Eq. (\ref{eq:Boltzmann}) corresponds to the laser absorption mechanisms in the CB. The third term describes the relaxation of electrons (no photon involved), and the last term describes the electron recombination to the valence band or states located in the bandgap, with a characteristic time $\tau_r$. Except the photo-ionization and the MIP, the collision integrals describing each process $P$ are calculated as follows \cite{Daguzan}:
\begin{equation}
\begin{array}{ll}
 \left. \frac{\partial f {(E_k)}}{\partial t} \right|_{P} = & \frac{V}{(2 \pi)^3} \left( \int W{_P (\vec k',\vec k)} f{(\vec k',t)} (1 - f{(\vec k,t)}) d\vec k' \right. \\
& \left. - \int W{_P (\vec k,\vec k')} f{(\vec k,t)} (1 - f{(\vec k',t)}) d \vec k' \right)
\end{array}
 \label{eq:collop}
\end{equation}
where $E_k = \hbar^2 k^2 / 2m$, $V$ is the volume of a lattice cell, and the rate $W_P$ is calculated according to the Fermi's golden rule. It mainly relies on the evaluation of a matrix element accounting for the coupling between an initial and a final state. The expression (\ref{eq:collop}) accounts for both the filling and the emptying of a given state of energy $E_k$, allowing one to conserve the density of conduction electrons after any transition in the CB. A detailed description of the presently used collision integrals can be found in \cite{PhysRevB.61.11437, Daguzan}; hereafter are discussed their main physical properties. Eq. (\ref{eq:Boltzmann}) is numerically solved by using an explicit scheme for time discretization on a fixed energy mesh.

In a strong laser field, the valence electrons are firstly promoted into the CB through the photo-ionization (PI) processes. At a low intensity ($\sim$ TW.cm\textsuperscript{-2}), the multiphoton absorption is dominant, i.e. several photons may be simultaneously absorbed. For higher intensities, tunnel ionization becomes dominant. In order to account for both multiphoton absorption and tunelling, the complete Keldysh expression \cite{JETP.20.1307, Gruzdev} is used to model the photo-ionization \cite{PhysRevB.61.11437}. The impact ionization, which may lead to electron avalanche, is not introduced here since it is negligible for laser pulse duration shorter than roughly 100 fs \cite{PhysRevB.61.11437}. Recent experimental investigations using time-resolved interferometry in fused silica confirm this statement \cite{Appl.Phys.A.110.709}. Our choice not to include impact ionization is also motivated by keeping the simplest reliable modeling to clearly exhibit the role of the MIP on the electron dynamics.

Three contributions are included to account for the laser driven excitation of electrons in the CB, i.e. the laser heating of the conduction electrons. The collisional operator for this process can then be split into 3 terms and thus reads:
\begin{equation}
  \begin{array}{lclcl}
    \left. \frac{\partial f(E_k, t)}{\partial t} \right|_{{Heat}}
    & = & \left. \frac{\partial f}{\partial t} \right|_{{e-ph-pt}}
    & + & \left. \frac{\partial f}{\partial t} \right|_{{e-i-pt}}
    +   \left. \frac{\partial f}{\partial t} \right|_{{MIP}}
  \end{array}
\label{eq:heating}
\end{equation}
where the first and second terms account for phonon-assisted and ion-assisted simultaneous absorption or emission of photons by electrons, respectively. They correspond to three-particle collisions, hereafter referred to as e-ph-pt and e-i-pt collisions, respectively. These processes include the possibility for electrons to absorb simultaneously several photons whatever their kinetic energy. Note that the e-i-pt process is nothing but the inverse bremsstrahlung mechanism \cite{PhysRevA.7.1064}. The last term of Eq. (\ref{eq:heating}) corresponds to the MIP which rate has been provided in the previous Section. Despite this process is described through a complex band structure, the transition rate provides a prediction only depending on the final electron energy $E_k$. We may thus adopt an \textit{ad hoc} procedure to include it in the kinetic approach by simply considering the contribution of this process as an usual rate $W_P$ as it appears in Eq. (\ref{eq:collop}). The multiphoton interband contribution in (\ref{eq:Boltzmann}) thus reads:
\begin{equation}
  \begin{array}{ll}
    \left. \frac{\partial f {(E_k)}}{\partial t} \right|_{{MIP}} = & \frac{V}{(2 \pi)^3} \left( \int_0^\infty \frac{\partial w_{1f}}{\partial k'} f{(k',t)} (1 - f{(k,t)}) k'^2 dk' \right. \\
& \left. - \int_0^\infty \frac{\partial w_{1f}}{\partial k'} f{(k,t)} (1 - f{(k',t)}) k'^2 dk' \right)
\end{array}
\end{equation}
with $E_k = \hbar^2 k^2 / 2m$.

The relaxation processes in Eq. (\ref{eq:Boltzmann}) are related to electron-electron (e-e) and electron-phonon (e-ph) interactions \cite{PhysRevB.61.11437, Daguzan}. These processes lead to the electron energy exchange from one to another or to the lattice, respectively. Here the electron-ion interaction for relaxation is neglected due to the low ionization degree within the present laser parameters. In the case of e-e interaction, the total energy of the electron gas is conserved. Note that the characteristic time of energy transfer depends on the electron density in the CB and may be as long as tens of fs in dielectric materials due to a relatively low produced electron density in the CB for moderate intensities lower than the ablation threshold \cite{PhysRevB.61.11437, Daguzan}. Due to e-ph interactions, the total energy of the free electron gas may decrease, leading to an increase in the phonon population. In our model, only optical phonon modes are included with energies of 63 meV and 153 meV. Their distribution is set according to the equilibrium Bose-Einstein statistics at room temperature. Note that the contribution of acoustic phonons to the energy transfer from electrons to the lattice is assumed to be negligible since the energy of acoustic phonons is significantly lower than the one of optical phonons \cite{Daguzan}.

Finally, in order to calculate the electron density for a given energy, the electron distribution function $f(E_k)$ is weighted by the density of states $g(E_k)$. The latter is assumed to evolve as $\sqrt{2 E_k m_e^3} / (\pi^2 \hbar^3)$ accounting for a three-dimensional free electron gas.

%


\section{Results and Discussion}
\label{section3}
Here we consider a material with a bandgap of 9 eV. It is representative of silica or quartz for the crystalline phase. The laser pulse has a gaussian shape with a full width at half-maximum of 70 fs, an intensity in the range of tens of TW/cm$^{2}$, and a wavelength of 800 nm. These parameters correspond to current laser facilities. Within these conditions, the absorption of 6 photons ($\hbar \omega$ = 1.55 eV) is required to bridge the bandgap in the multiphoton regime.
\begin{figure}[t]
  \begin{center}
    \subfigure[\label{Figure_2_a}]{\includegraphics[width = 10cm, trim = 0cm 0cm 0cm 0cm, clip] {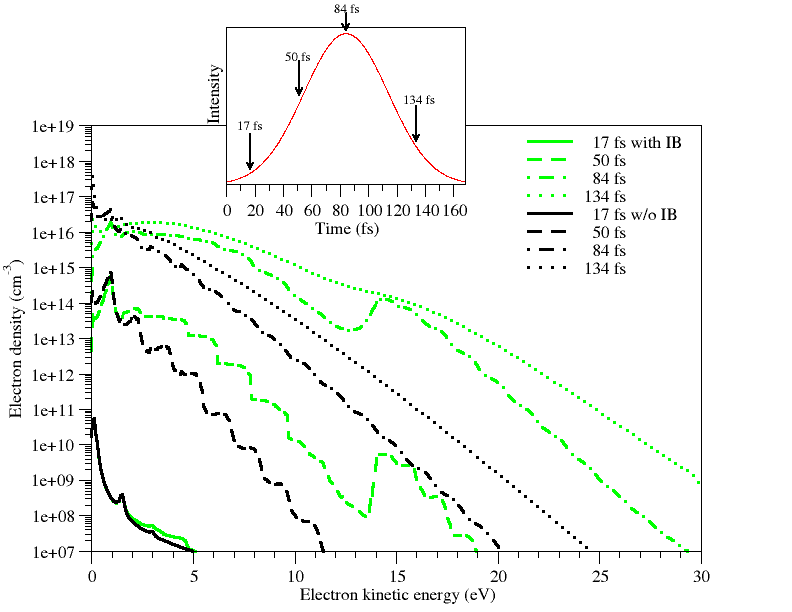}} \quad
    \subfigure[\label{Figure_2_b}]{\includegraphics[width = 9.5cm, trim = 0cm 0cm 0cm 3cm, clip] {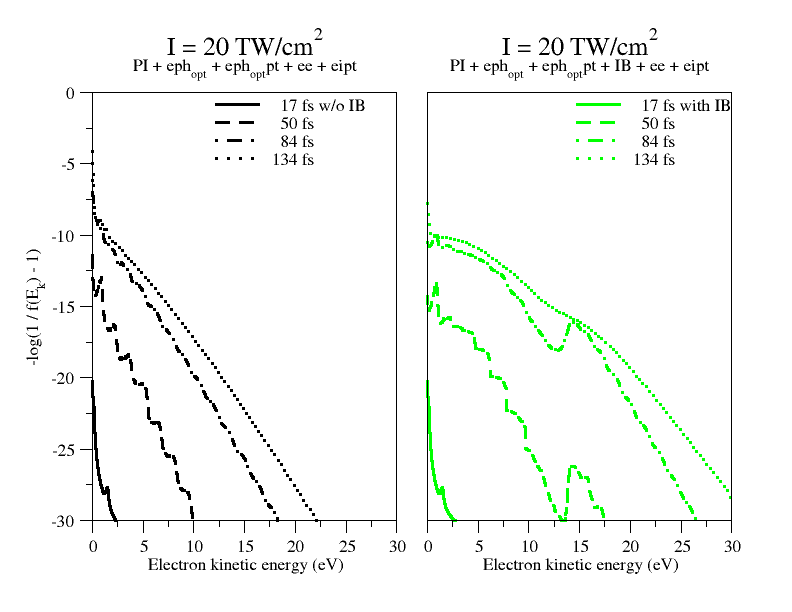}}
  \end{center}
  \caption{(a) Evolution of the electron density as a function of the electron kinetic energy for various times during the laser irradiation. The model includes the multiphoton interband process or not (green and black color respectively). The temporal evolution of the laser intensity is plotted to illustrate the studied times. (b) Same as previously but with the function $- \ln (1/f(E_k) - 1)$ ((left) without MIP; (right) with MIP). Within this representation, a straight line corresponds to the Fermi-Dirac distribution. The maximum laser intensity is set to 20 TW/cm$^{2}$ and the FWHM pulse duration is 70 fs.}
  \label{Figure_2}
\end{figure}

As a preliminary comment regarding the electron dynamics with the present laser parameters, the recombination of conduction electrons is neglected due to the short interaction time compared to the characteristic recombination time of 150 fs \cite{Martin}. In addition, since the recombination time does not depend on the electron energy, it may only affect the total electron density but not the shape of the electron energy distribution. A weak influence of the recombination time on the forthcoming numerical results has been confirmed. We emphasize that the removing of non contributing processes in dedicated simulations allows us to better highlight the influence of the MIP.

Figure \ref{Figure_2_a} shows the electron density in the CB as a function of the electron energy for four times during the interaction: 17 fs, 50fs, 84fs, and 134fs. The inset depicts these times within the temporal pulse envelope. In that case, the maximum intensity is set to 20 TW/cm$^{2}$. In order to evaluate the influence of the MIP, the electron energy distributions are calculated by using the above-presented model by including or not the multiphoton interband transitions (green or black curves, respectively). For the first studied time of 17 fs, which corresponds to the beginning of the interaction (solid curves), both distributions (with or without the MIP) are similar, and exhibit 2 peaks. The first peak close to the threshold corresponds to electrons directly promoted from the valence band through the simultaneous absorption of 6 photons. This first peak is followed by another one centered around 1.6 eV. It corresponds to the further absorption of one photon in the CB due to the e-ph-pt interaction. For this short time where the intensity remains very moderate (of the order of 1 TW/cm$^{2}$), an examination of the MIP rate shows that it is significantly smaller than the e-ph-pt rate, explaining the similar appearance of both distributions.
\begin{figure}[t] 
  \begin{center}
    \includegraphics[width = 10cm, trim = 0cm 0cm 0cm 0cm, clip] {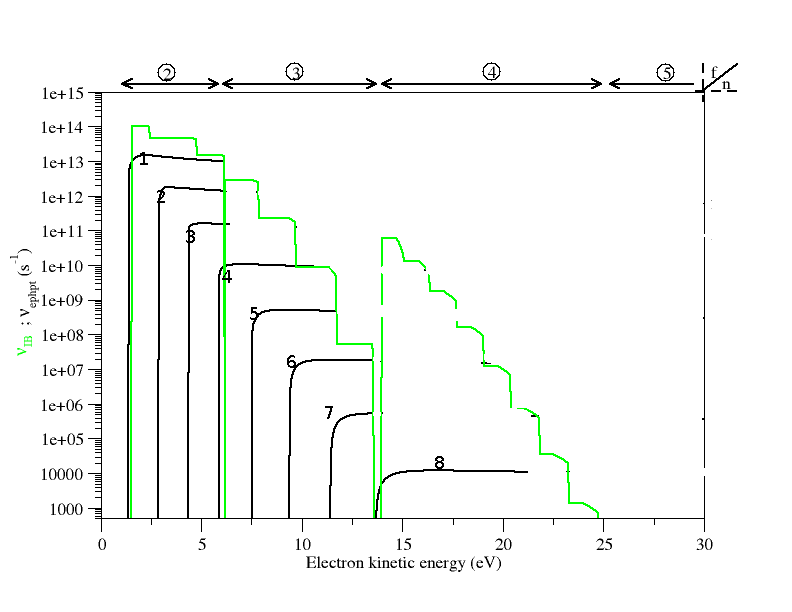}
  \end{center}
  \caption{Evolution of the multiphoton interband rate (green line) and the e-ph-pt rate (black lines) as a function of the final electron energy. The laser intensity is set to 10 TW/cm$^2$.} 
  \label{fig:interband_rate}
\end{figure}

When the time is elapsing up to 50 fs (dashed curves), the total electron density increases and the distributions exhibit the same shape as the previous time but including more peaks. The latter are still separated by the photon energy, and their large number is due to more efficient photon absorption processes in the CB when the laser intensity is growing. Now, the predictions of the model including interband transitions significantly depart from the other. In that case, the contribution of direct vertical transitions in the Brillouin zone becomes significant leading to a population of electrons with higher energies. In order to better understand this behavior, the evolution of the MIP and e-ph-pt rates (term $W_P$ appearing in Eq. (\ref{eq:collop})) are plotted as a function of the electron energy in Fig. \ref{fig:interband_rate} for I = 10 TW/cm$^2$. Regarding the e-ph-pt rate, its value for each number of absorbed photons (from 1 to 8) is provided and has been plotted only to energy values consistent with the number of involved photons for an initial energy relevant of the first band. This case is comparable to the MIP where a state with energy $E_k$ is filled from electrons located in the bottom of the CB (first band). It appears clearly that the MIP rate is higher than the e-ph-pt rate in all conditions. For the e-ph-pt process, note that a state with energy $E_k$ may be filled by electrons with energy $E_k - n\hbar \omega$ whatever $n$. For instance, electrons with 21.55 eV may be generated from electrons of energy 20 eV through a one-photon absorption process. However, since the energy distribution decreases exponentially as a function of the electron energy, the contribution of such processes is also negligible compared to the MIP. The higher MIP rate compared to the e-ph-pt rate is explained by the fact that they involve 2 and 3 particles interaction, respectively: the higher the number of interacting particles, the lower the interaction probability.


The energy distribution in Fig. \ref{Figure_2_a} including the MIP also exhibits a particular feature around 14eV where an unexpected increase in the distribution takes place. As shown in Fig. \ref{fig:interband_rate}, this behavior is a direct consequence of the evolution of the MIP rate with respect to the energy: such a non monotonic evolution is observed at the transition from the third to the fourth band. This is due to the non perturbative behavior of the multiphoton rate in conditions of large momentum transfer (see argument of the Bessel function in Eq. (\ref{eq:B1f})).
\begin{figure}[t]
  \begin{center}
    \includegraphics[width = 10cm, trim = 0cm 0cm 0cm 3cm, clip] {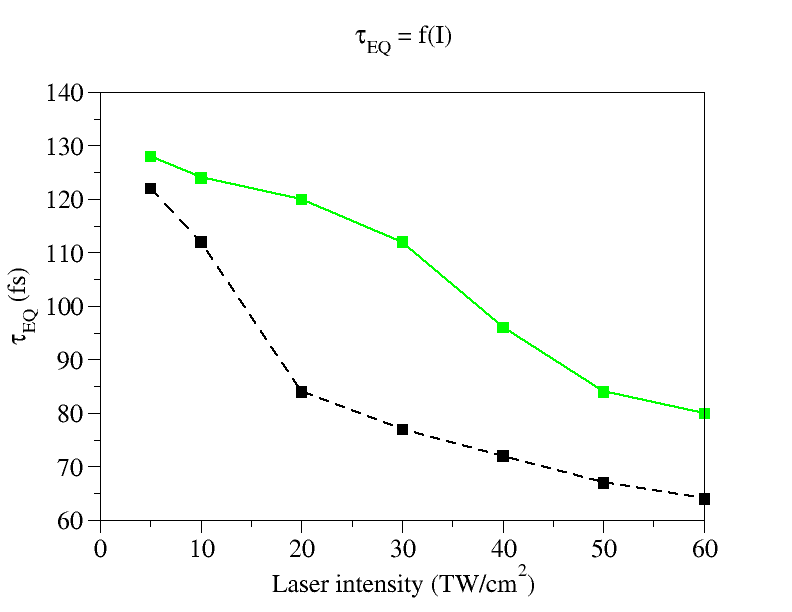}
  \end{center}
  \caption{Evolution of the equilibrium time $\tau_{_{EQ}}$ as a function of the maximum laser intensity for a 70 fs gaussian pulse. The model includes or not the multiphoton interband process (green solid line or black dashed line, respectively).}
  \label{Figure_3}
\end{figure}

At the peak intensity corresponding to the time of 84 fs, the appearance of the distribution function changes dramatically due to the influence of electron-electron and electron-phonon collisions which induce a relaxation of the electron gas. For this long enough interaction time, a significant energy exchange gives rise to a smoothing of the electron energy distribution, in particular of the photon absorption imprint. In the case where MIP is switched off, the distribution is close to a Fermi-Dirac distribution accounting for an equilibrium state. This is demonstrated in Fig. \ref{Figure_2_b} which shows the function $- \ln (1/f(E_k) - 1)$ as a function of the electron kinetic energy for the times under consideration. Indeed, a linear behavior of this function corresponds to the equilibrium state \cite{PhysRevB.61.11437}. When the MIP is allowed, the equilibrium is reached after a longer time close to 130 fs as shown in Fig. \ref{Figure_2_b}. Since the MIP heats more efficiently the electrons, more time is required for the electron gas to relax towards an equilibrium distribution.

This behavior is further highlighted in Fig. \ref{Figure_3} that shows the evolution of the equilibrium time $\tau_{_{EQ}}$ as a function of the maximum laser intensity, with and without the MIP (solid and dashed lines, respectively). $\tau_{_{EQ}}$ is defined as the time needed for the electron energy distribution to reach the equilibrated Fermi-Dirac distribution within an accuracy of 5 \% (root mean square). Whatever the laser intensity, this figure confirms an increase in $\tau_{_{EQ}}$ due to the introduction of the MIP which leads electrons to higher energies. Due to the non-monotonic evolution of the MIP rate with the intensity, the relative increase of $\tau_{_{EQ}}$ also depends on the intensity and can reach up to 40 \% for 20 TW/cm$^2$. Now, whatever the introduced heating processes, the overall behavior of $\tau_{_{EQ}}$ with respect to the intensity is the same: the higher the intensity, the shorter the relaxation time. Following the previous considerations, the opposite behavior would have been expected since higher intensities lead to higher electron energies. Actually, the increase in the intensity also leads to an increase in the electron density, thus decreasing the characteristic time for e-e and e-ph collisional relaxation processes. Since the electron production in the CB is a highly nonlinear process (at least 6 photons are required to bridge the band gap) with respect to the intensity, it turns out that its contribution to the value of $\tau_{_{EQ}}$ is larger than the one of the heating processes in the CB. Despite the MIP is itself nonlinear, it includes lower order photon-absorption processes, which lead to moderate variations with respect to the intensity as long as the latter is not too high. Note that for the range of intensities under consideration, the produced electron density in the CB is of the order of 10$^{19}$-10$^{20}$ cm$^{-3}$.
\begin{figure}[t]
  \begin{center}
    \includegraphics[width = 10cm, trim = 0cm 0cm 0cm 3cm, clip] {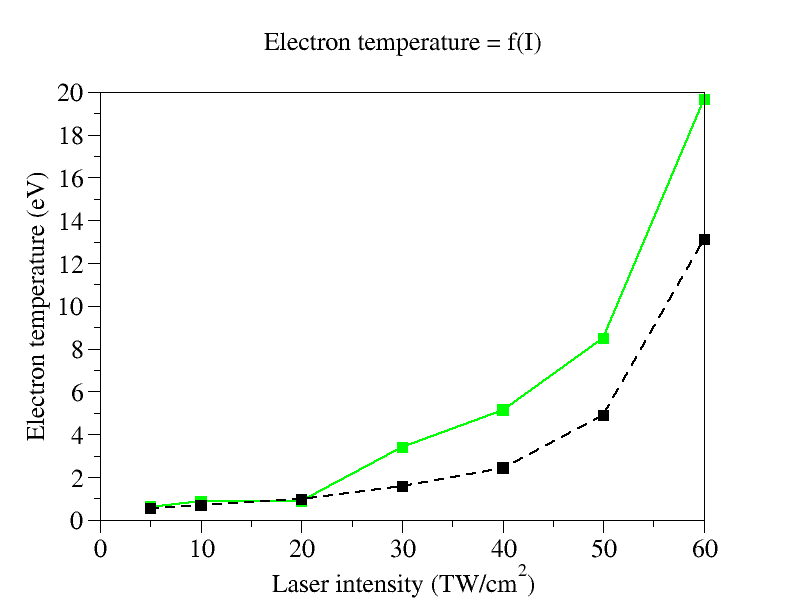}
  \end{center}
  \caption{Evolution of the electron temperature as a function of the laser intensity at the end of the laser pulse ($t=$190 fs). The model includes or not the multiphoton interband process (green solid line or black dashed line respectively.}
  \label{fig:temperature}
\end{figure}

When the equilibrium is reached, a temperature of the electron gas can be defined. Figure \ref{fig:temperature} shows this temperature just after the interaction ($t=$190 fs) as a function of the laser intensity with and without the MIP. Both curves exhibit a nonlinear increase with respect to the intensity, accounting for the multiphoton absorption in the conduction band. Regarding the additional influence of the MIP on the overall heating of the electron gas, it becomes significant for intensities above 20 TW/cm$^2$. For smaller intensities, despite a fraction of electrons may be promoted to higher energies, their contribution is not large enough to significantly modify the total energy of the electron gas. Indeed, despite the MIP rate is higher than the e-ph-pt rate for these intensities (see for instance Fig. \ref{fig:interband_rate} for $I=$10 TW/cm$^2$), it only provides energy to electrons in the bottom of the CB (first band) whereas electrons may absorb one or more photons through the e-ph-pt process whatever their energy.
\begin{figure}[t]
  \begin{center}
    \includegraphics[width = 10cm, trim = 1cm 0cm 0cm 3cm, clip] {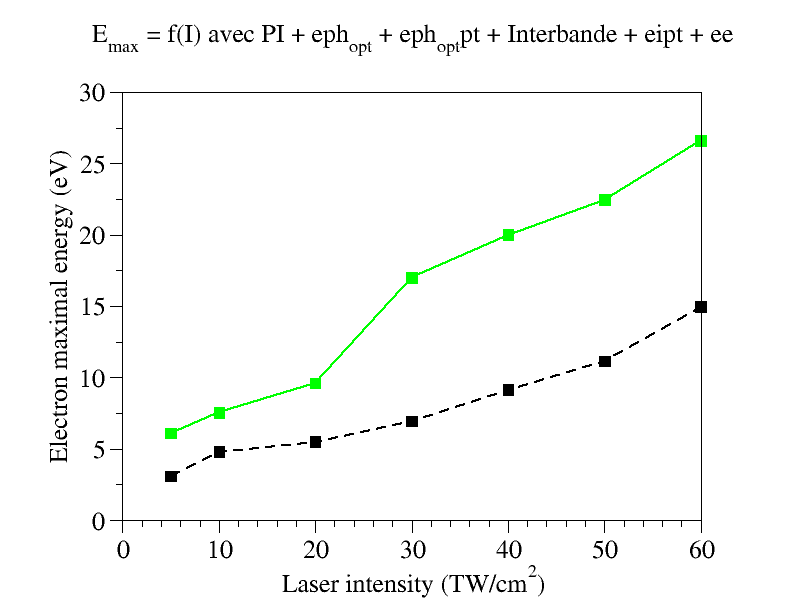}
  \end{center}
  \caption{(color online) Evolution of the electron maximal energy as a function of the laser intensity. This value is obtained at half the pulse duration. The model includes or not the multiphoton interband process (green solid line and black dashed line, respectively.)}
  \label{Figure_4}
\end{figure}

Finally, Fig. \ref{Figure_4} shows the electron maximum energy, $E_{max}$, at half the pulse duration (when relaxation processes are not yet important) as a function of the laser intensity. This energy is evaluated as follows. For a given intensity, the electron energy distribution is first divided by the total electron density to avoid any influence of the electron production. $E_{max}$ is defined as the value of the normalized distribution at the level of 0.5 \% of its maximum value. Note that small variations of the chosen level lead to similar conclusions. As expected, Fig. \ref{Figure_4} shows that the higher the intensity, the larger the maximum electron energy whatever the considered laser excitation process in the CB. The MIP also increases the maximum energy. This increase appears not to depend significantly on the intensity, with a mean value of the order of 100 \%, i.e. roughly a factor 2. As for the relaxation time, variations of the influence of the MIP heating with respect to the intensity are observed, accounting for the non monotonic behavior of the interband rate with respect to the electron energy. Despite the MIP is nonlinear with respect to intensity, thus providing a nonlinear evolution of $E_{max}$ with respect to the intensity, the increase in intensity also gives rise to a higher electron density and subsequently more effective relaxation processes. In particular, electron-electron collisions, which redistribute the electron energies, lead to lower maximum energy than with a pure MIP heating. In addition, the same argument as for the equilibration time explains the relatively slow variations of $E_{max}$ with respect to the intensity. It is close to a linear behavior.

\section{Conclusions}
\label{section4}
In the case of irradiation of dielectric materials by femtosecond laser pulses of moderate intensities of a few TW/cm$^{2}$, it has been suggested in \cite{Appl.Phys.A.98.679, PhysRevB.74.235215, J.PhysiqueIV.108.113, Europhys.Lett.67.301, Hawkins2015} that direct multiphoton transitions between sub-bands of the conduction band may contribute to the electron dynamics, in particular to higher electron energies. The present work deals with the case of higher intensities in the range of tens of TW/cm$^{2}$. The expression for the multiphoton interband rate has been revisited and generalized to highly excited sub-bands of the conduction band. This non-collisional heating process has been introduced for the first time in a Boltzmann kinetic equation including the standard collisional heating and relaxation interactions. The way to introduce consistenly this non-collisional process in the kinetic approach has also been provided. Under this framework, the electron dynamics in a wide bandgap dielectric material has been investigated. The multiphoton interband process has been shown to significantly modify the energy distribution of electrons compared to the standard collisional processes. The maximum electron energy is increased by roughly a factor two. That leads to a longer time for the electron gas to relax towards an equilibrium state. Calculations show this relative increase of the relaxation time may be of the order of 40 \% depending on the laser intensity. The main conclusion of the present work is that the overall electron dynamics is significantly modified by direct multiphoton interband transitions. This mechanism thus should be taken into account in future developments aiming at describing the electron dynamics in dielectric materials to predict local modifications and possible applications as described in the introduction. Based on the proposed framework, more accurate predictions could also be performed by improving the description of the laser-induced electron transitions in the conduction band.

The present model provides predictions which can be compared to electron energy distributions obtained from photo-emission experiments \cite{Geoffroy, Fedorov}. Such a comparison is in progress for laser intensities lower than the ablation threshold. Preliminary results show that the calculated evolution of the maximum electron energy with respect to the intensity is in a relatively good agreement with experimental results up to energies in excess of 50 eV. This demonstrates the reliability of the present modeling and will be addressed in details in a forthcoming work.


\section*{Acknowledments}
Henri Bachau is acknowledged for fruitul discussions and for providing a code solving the time-dependent Schrodinger equation. We acknowledge the financial support (Ph'D grant) of both the Commissariat à l'énergie atomique (CEA) and the Conseil Régional d'Aquitaine. The authors thank the University of Bordeaux for providing access to the Mésocentre de Calcul Intensif Aquitain (MCIA).

\appendix
\section{Derivation of the multiphoton interband rate}
\label{app:interband}
In the presence of an external laser electric field, conduction electrons may be excited from a sub-band to another one through multiphoton absorption. Here is derived the expression for the direct multiphoton transition between the first lowest and higher excited sub-band of the conduction band based on previous works \cite{NIMPRB.107.165, Europhys.Lett.67.301, Appl.Phys.B.78.989, Appl.Phys.A.98.679, Phys.Status.Solidi.C.2.240}.

The rate per unit volume for the transition from an initial sub-band (1) to a final sub-band (f) is given by:
\begin{equation}
  \begin{array}{lcl}
    w_{f1} & = & \frac{2 \pi}{\hbar \left( 2 \pi \hbar \right)^3} \sum\limits_{n}
    \int \left| M_{f1}(\vec{p}) \right|^2 
    \delta ( \bar{E} (\vec{p}(t), \vec{F}) - n \hbar \omega) d\vec{p}
  \end{array}
  \label{eq:depart}
\end{equation}
where $M_{f1}(\vec{p})$ is the matrix element for the considered transition, which depends on the momentum $\vec{p}$ of the first Brillouin zone. The matrix element reads \cite{NIMPRB.107.165, Europhys.Lett.67.301, Appl.Phys.B.78.989, Appl.Phys.A.98.679, Phys.Status.Solidi.C.2.240}:
\begin{equation}
  \begin{array}{lcl}
    M_{f1}(\vec{p}) = \frac{1}{2 \pi} \oint V_{1f} (\vec{p}(t)) e^{\frac{i}{\hbar \omega}
      \int\limits_{0}^{u} E(\vec{p}(t), \vec{F}) \frac{d\nu}{\sqrt{1 - \nu^2}} } du
  \end{array}
  \label{eq:matrixelement}
\end{equation}
The Dirac delta function provides selection rules for the absorption of $n$ photons, accounting for the energy conservation. $\bar{E} (\vec{p}(t), \vec{F})$ is the average over the laser field phase of the time-dependent energy gap between the initial and final sub-bands, reading $E(\vec{p}(t), \vec{F}) = E_f (\vec{p}(t), \vec{F}) - E_1 (\vec{p}(t), \vec{F})$ where $E_i (\vec{p}(t), \vec{F}) = \frac{\left( \vec{p}(t) - \vec{p}_i \right)^2}{2 m}$ is the electron energy of quasi-momentum $\vec{p}$ in the band $i$. $\vec{p}_i$ is a quasi-momentum of the reciprocal-lattice vector. For a cubic structure, $\vec{p}_i$ is given by $\hbar \frac{2 \pi j_i}{a} \vec{e_{p_i}}$ with $a$ the lattice period and $j_1 = 0$, $j_2 = 1$, $j_3 = - 1$, $j_4 = 2$, $j_5 = - 2$, etc. The time-dependent momentum reads $\vec{p}(t) = \vec{p} + e \frac{\vec{F}}{\omega} sin(\omega t)$ where $\vec{F}$ is the envelope of the laser electric field. It follows that:
\begin{equation}
  \begin{array}{lcl}
    \bar{E} (\vec{p}(t), \vec{F})
    & = & \frac{1}{2 \pi} \int\limits_{-\pi}^{+\pi} E(\vec{p}(x/\omega), \vec{F}) dx \\
    & = & \frac{\vec{p} . \left( \vec{p}_1 - \vec{p}_f \right)}{m} + \frac{p_f^2 - p_1^2}{2 m}
  \end{array}
\end{equation}
Using a cubic lattice structure, $\vec{p}_1 = \vec{0}$ and considering that $\vec{p} . \vec{p}_f = p {p}_f \cos\theta$ then
\begin{equation}
  \begin{array}{lcl}
    \delta ( \bar{E} (\vec{p}(t), \vec{F}) - n \hbar \omega)
    = \frac{m}{\left| p_f \cos\theta \right|} \delta ( p - \frac{p_n}{\cos\theta}) \ , \\
  \end{array}
\end{equation}
where $p_n = \sqrt{2 m E_n} = \frac{m}{p_f} (\frac{p_f^2}{2 m} - n \hbar \omega)$. That permits to identify the energy $E_n$:
\begin{equation}
  \begin{array}{lcl}
    E_n & = &
    \left\{ \begin{array}{lcl} 
      \left( 1 - \frac{n}{\tilde{n}} \right)^2 E_{0}	
      & \mbox{if} & n \in \left[ 1 ; \left< \tilde{n} \right> \right] \mbox{or } n \in \left] \left< \tilde{n} \right> ; \left< 2 \tilde{n} \right> \right] \\
	  4 \left( 1 - \frac{n}{4 \tilde{n}} \right)^2 E_{0}
	  & \mbox{if} & n \in \left] \left< 2 \tilde{n} \right> ; \left< 4 \tilde{n} \right> \right] \mbox{or } n \in \left] \left< 4 \tilde{n} \right> ; \left< 6 \tilde{n} \right> \right] \\
		  ... & \\
    \end{array} \right.
  \end{array}
\end{equation}
with $\tilde{n} = \frac{4 E_{0}}{\hbar \omega}$, $E_{0} = \frac{\pi^2 \hbar^2}{2 m a^2}$ being the limit of the first Brillouin zone of first sub-band. The different domains correspond to the transition to the sub-band 2, sub-band 3, etc.

By using the spherical coordinates to perform the integration over the momentum, Eq. (\ref{eq:depart}) then reads:
\begin{equation}
  w_{f1} = \frac{4 \pi^2}{\hbar \left( 2 \pi \hbar \right)^3} \sum\limits_{n}
  \int_0^{\pi/a} dp \ p^2 \int_0^\pi d\theta \sin \theta \left| M_{f1}(p,\cos \theta) \right|^2 
  \frac{m}{\left| p p_f \right| } \delta (\cos\theta - \frac{p_n}{p})
\end{equation}
where the coordinates have been oriented such that the integration over the azimuthal angle $\varphi$ is $2\pi$. By using the properties of the Dirac Delta function, the integration over $\theta$ leads to:
\begin{equation}
  w_{f1} = \frac{4 \pi^2}{\hbar \left( 2 \pi \hbar \right)^3} \sum\limits_{n}
  \int_0^{\pi/a} dp \ p^2 \left| M_{f1}(p,p_n/p) \right|^2 
  \frac{m}{\left| p p_f \right| } \theta(p - p_n)
\end{equation}
The last integration over the momentum may be performed by assuming conditions where the laser electric field is not too high, i.e. $eF/\omega \ll 1$, $V_{1f} (\vec{p}(t)) \simeq V_{1f} \sqrt{I}$ where $V_{1f}$ is a constant and $I$ is the laser intensity (related to the electric field as $I = n_0 \epsilon_0 c F^2 / 2$, with $n_0$ the index of refraction, $\epsilon_0$ the vacuum permittivity, and $c$ the speed of light in the vacuum). Since the wavefunction are unknown is the general case, $V_{1f}$ may not be easily determined and is the only free parameter of the present analytical approach.

Now, the argument of the exponential function appearing in Eq. (\ref{eq:matrixelement}) may be calculated by using the variable change $\nu = \sin(x)$. By using the above-given definition of $p_n$, defining $B_{1f} = \frac{1}{\hbar \omega} \frac{e \vec{F} \left( \vec{p}_f - \vec{p}_1 \right)}{m \omega}$, and with $n'=n$, one gets:
\begin{equation}
    M_{f1}(p,p_n/p) = M_{f1}(n) = V_{1f} \sqrt{I} \frac{1}{2 \pi} e^{- i B_{1f}} \oint\limits \cos(x) e^{- i \left( n' x - B_{1f} \cos(x) \right)} dx
\label{eq:matrixelement2}
\end{equation}
The previous expression includes the derivative of the Bessel function with respect to $B_{1f}$. Eq. (\ref{eq:matrixelement2}) then transforms into:
\begin{equation}
  M_{f1}(n) = - i V_{1f} \sqrt{I} J'_{n'}(B_{1f}) e^{i n \frac{\pi}{2}} e^{- i B_{1f}} \\
\end{equation}
The squared modulus thus can be written as:
\begin{equation}
  \left| M_{f1}(n) \right|^2 = V_{1f}^2 I J'^2_{n}(B_{1f})
\end{equation}
where $J'^2_{n'}(B_{1f}) = J'^2_{n}(B_{1f})$ has been used. By using the relation $J'_n(B_{1f}) = \frac{n}{B_{1f}} J_n(B_{1f}) - J_{n+1}(B_{1f})$, another form reads:
\begin{equation}
  \left| M_{f1}(n) \right|^2 = \frac{V_{1f}^2 I}{B_{1f}^2} \left( B_{1f} J_{n+1}(B_{1f}) - n J_n(B_{1f}) \right)^2
\end{equation}
The remaining momentum integration is straightforward and the following expression for the rate per unit volume can be obtained:
\begin{equation} \label{Eq_w}
  w_{f1}
  = \frac{m}{8 \pi \hbar^2 \left| p_f \right|} V_{1f}^2 I \sum\limits_{n} 2 J'^2_n(B_{1f}) \left[ \frac{\pi^2}{a^2} - k_n^2 \right]
\end{equation}


\vspace{2cm}

\begin{thebibliography}{99}
\bibitem{PhysRevB.65.214303}
B.~Rethfeld, A.~Kaiser, M.~Vicanek, and G.~Simon, {\em Phys. Rev. B}~{\bf 65}, pp.~214303(1)--214303(11), 2002.

\bibitem{PhysRevB.87.035414}
B.~Scharf, V.~Perebeinos, J.~Fabian, and P.~Avouris, {\em Phys. Rev. B}~{\bf 87}, pp.~035414(1)--035414(9), 2013.

\bibitem{PhysRevB.54.4660}
P.~E. Selbmann, M.~Gulia, F.~Rossi, and E.~Molinar, {\em Phys. Rev. B}~{\bf 54}, pp.~4660--4673, 1996.

\bibitem{PhysRevB.61.11437}
A.~Kaiser, B.~Rethfeld, M.~Vicanek, and G.~Simon, {\em Phys. Rev. B}~{\bf 61}, pp.~11437--11449, 2000.

\bibitem{Appl.Phys.A.110.579}
N.~Shcheblanov and T.~Itina, {\em Appl. Phys. A}~{\bf 110}, pp.~579--583, 2013.

\bibitem{Bulgakova2006}
S.~Winkler, I.~Burakov, R.~Stoian, N.~Bulgakova, A.~Husakou,
  A.~Mermillod-Blondin, A.~Rosenfeld, D.~Ashkenasi, and I.~Hertel {\em Appl.
  Phys. A}~{\bf 84}, p.~413, 2006.

\bibitem{Bulgakova2010}
N.~Bulgakova, R.~Stoian, and A.~Rosenfeld {\em Quantum Electron.}~{\bf 40},
  p.~966, 2010.

\bibitem{Rep.Prog.Phys.76.036502}
P.~Balling and J.~Schou, {\em Rep. Prog. Phys.}~{\bf 76}, pp.~036502(1)--036502(39), 2013.

\bibitem{PhysRevE.85.56403}
A.~Bourgeade and G.~Duchateau, {\em Phys. Rev. E}~{\bf 85}, p.~056403, 2012.

\bibitem{DuchateauBourgeade}
G.~Duchateau and A.~Bourgeade {\em Phys. Rev. A}~{\bf 89}, p.~053837, 2014.

\bibitem{GulleySPIE}
J.~R. Gulley {\em Proc. SPIE}~{\bf 7842}, p.~78420U, 2010.

\bibitem{Gulley}
J.~R. Gulley and W.~M. Dennis {\em Phys. Rev. A}~{\bf 81}, p.~033818, 2010.

\bibitem{DuchateauPop}
G.~Duchateau, {\em Phys. Plasmas}~{\bf 20}, p.~022306, 2013.

\bibitem{Varin}
C. Varin, C. Peltz, T. Brabec, and T. Fennel, {\em Phys. Rev. Lett.} {\bf108}, 175007 (2012).

\bibitem{Otobe2008}
T.~Otobe, M.~Yamagiwa, J.-I. Iwata, K.~Yabana, T.~Nakatsukasa, and G.~Bertsch, {\em Phys. Rev. B}~{\bf 77}(16), p.~165104, 2008.

\bibitem{Otobe2010}
T.~Otobe, {\em J. Phys.: Condens. Matter}~{\bf 22}(38), p.~384204, 2010.

\bibitem{PhysRevB.77.165104}
T.~Otobe, M.~Yamagiwa, J.~I. Iwata, K.~Yabana, and T.~Nakatsukasa, {\em Phys. Rev. B}~{\bf 77}, pp.~1651404 (1--5), 2008.

\bibitem{J.Opt.Soc.Am.B.31.C28}
N.~Brouwer and B.~Rethfeld, {\em J. Opt. Soc. Am. B}~{\bf 31}, pp.~C28--C35, 2014.

\bibitem{Appl.Phys.A.98.679}
H.~Bachau {\it et al}, {\em Appl. Phys. A}~{\bf 98}, pp.~679--689, 2009.

\bibitem{PhysRevB.74.235215}
H.~Bachau, A.~N. Belsky, P.~Martin, A.~N. Vasil'ev, and B.~N. Yatsenko, {\em Phys. Rev. B}~{\bf 74}, p.~235215, 2006.

\bibitem{J.PhysiqueIV.108.113}
A.~Belsky {\it et al}, {\em J. Phys. IV}~{\bf 108}, pp.~113--117, 2003.

\bibitem{Europhys.Lett.67.301}
A.~Belsky {\it et al}, {\em Europhys. Lett.}~{\bf 67}, pp.~301--306, 2004.

\bibitem{Hawkins2015}
P.~Hawkins, M.~Ivanov, and V.~Yakovlev {\em Phys. Rev. A}~{\bf 91}, p.~013405,
  2015.

\bibitem{Bulgakova2004}
N.~Bulgakova, R.~Stoian, A.~Rosenfeld, I.~Hertel, and E.~Campbell {\em Phys.
  Rev. B}~{\bf 69}, p.~054102, 2004.

\bibitem{Bulgakova2016}
N.~Bulgakova, V.~Zhukov, I.~Mirza, Y.~Meshcheryakov, J.~Tomáštík,
  V.~Michálek, O.~Haderka, L.~Fekete, A.~Rubenchik, M.~Fedoruk, and T.~Mocek
  {\em Proc SPIE}~{\bf 9735}, p.~97350N, 2016.

\bibitem{Mezel}
C.~Mezel, G.~Duchateau, G.~Geneste, and B.~Siberchicot, {\em J. Phys.: Condens. Matter}~{\bf
  25}, p.~235501, 2013.

\bibitem{Appl.Phys.B.78.989}
A.~Belsky {\it et al}, {\em Appl. Phys. B}~{\bf 78}, pp.~989--994, 2004.

\bibitem{Phys.Status.Solidi.C.2.240}
B.~N. Yatsenko {\it et al}, {\em Phys. Status. Solidi. C}~{\bf 2}, pp.~240--243, 2005.

\bibitem{Shcheblanov}
N.~Shcheblanov and T.~Itina {\em Applied Physics A}~{\bf 110}, p.~579, 2012.

\bibitem{Ashcroft}
N.~W. Ashcroft and N.~D. Mermin, {\em Solid State Physics}, EDP Sciences, 1976.

\bibitem{JETP.20.1307}
L.~V. Keldysh, {\em Sov. Phys. JETP}~{\bf 20}, pp.~1307--1314, 1965.

\bibitem{Gruzdev}
V.~Gruzdev {\em Phys. Rev. B}~{\bf 75}, p.~205106, 2007.

\bibitem{Volkov}
D.~M. Volkov {\em Z. Phys.}~{\bf 94}, p.~250, 1935.

\bibitem{Ridley}
B.~Ridley, {\em Quantum Processes in Semiconductors}, Oxford University Press,
  1999.

\bibitem{DuchateauThese}
G.~Duchateau, E.~Cormier, and R.~Gayet {\em Phys. Rev. A}~{\bf 66}, p.~023412,
  2002.

\bibitem{Rodriguez}
V.~Rodriguez, E.~Cormier, and R.~Gayet {\em Phys. Rev. A}~{\bf 69}, p.~053402,
  2004.

\bibitem{Catoire}
F.~Catoire and H.~Bachau {\em Phys. Rev. Lett.}~{\bf 115}, p.~163602, 2015.

\bibitem{Daguzan}
P.~Daguzan, P.~Martin, S.~Guizard, and G.~Petite {\em Phys. Rev. B}~{\bf 52},
  p.~17099, 1995.

\bibitem{Appl.Phys.A.110.709}
A.~Mouskeftaras, S.~Guizard, N.~Fedorov, and S.~Klimetov, {\em Appl. Phys. A.}~{\bf 110}, pp.~709--715, 2013.

\bibitem{PhysRevA.7.1064}
J.~F. Seely and E.~G. Harris, {\em Phys. Rev. A}~{\bf 7}, pp.~1064--1067, 1973.

\bibitem{Martin}
P.~Martin, S.~Guizard, P.~Daguzan, G.~Petite, P.~D'Oliveira, P.~Meynadier, and
  M.~Perdrix {\em Phys. Rev. B}~{\bf 55}, p.~5799, 1997.

\bibitem{Geoffroy}
G.~Geoffroy, G.~Duchateau, N.~Fedorov, P.~Martin, and S.~Guizard {\em Laser
  Physics}~{\bf 24}, p.~086101, 2014.

\bibitem{Fedorov}
N. Fedorov {\it et al}, {\em J. Phys.: Condens. Matter} {\bf 28}, 315301, 2016.

\bibitem{NIMPRB.107.165}
A.~Vasil'ev, {\em NIMPR. B}~{\bf 107}, pp.~165--171, 1996.
\end{thebibliography}
\end{document}